\documentclass[aps,prb,preprint,groupedaddress,citeautoscript,nopacs]{revtex4}
\usepackage{graphicx}
\usepackage{latexsym}
\usepackage{amsmath}
\usepackage{amsthm}
\usepackage{amssymb}
\usepackage{epstopdf} 
\usepackage{enumerate}
\usepackage{setspace} 
\usepackage{dcolumn}
\usepackage{bm}
\usepackage{setspace} 
\usepackage{slashed}
\usepackage{color}
\usepackage{youngtab}

\newcommand\Tr{\mathop{\mathrm{Tr}}}

\begin{document}
\title{
Competing Orders and Anomalies 
}

\author{Eun-Gook Moon }

\affiliation{ Kadanoff Center for Theoretical Physics, University of Chicago, Chicago, IL 60637, USA}

\affiliation{Department of Physics, Korea Advanced Institute of Science and Technology, Daejeon 305-701, Korea}

\date{\today}

\begin{abstract} 
A conservation law  is one of the most fundamental properties in nature, but a certain class of  conservation ``laws'' could be spoiled by intrinsic quantum mechanical effects, so-called quantum anomalies. 
Profound properties of the anomalies have deepened our understanding in quantum many body systems.   
Here, we investigate quantum anomaly effects in quantum phase transitions between competing orders and  striking consequences of their presence. We explicitly calculate topological nature of anomalies of non-linear sigma models (NLSMs) with the Wess-Zumino-Witten (WZW) terms. The non-perturbative nature is directly related with the 't Hooft anomaly matching condition : anomalies are conserved in renormalization group flow. By applying the matching condition, 
we show massless excitations are enforced by the anomalies in a whole phase diagram in sharp contrast to the case of the Landau-Ginzburg-Wilson theory which only has massive excitations in symmetric phases.
Furthermore, we find non-perturbative criteria to characterize  quantum phase transitions between competing orders. For example, in $4D$, we show the two competing order parameter theories, $CP(1)$ and the NLSM with WZW, describe different universality class. Physical realizations and experimental implication of the anomalies are also discussed.   
\end{abstract}

\maketitle

{\textbf{\large Introduction}} 

Quantum anomaly is one of the most fascinating phenomena in quantum many body systems.
Quantum fluctuations spoil classical symmetry, thus corresponding conservation laws and Ward identities no longer hold and must be modified. Topological source terms to the ``conservation'' laws are induced by quantum fluctuations from anomalies, thus topology and symmetry are intrinsically tied.
Consequences of the anomalies were first confirmed in the pion-decay ($\pi^0 \rightarrow \gamma \gamma$), and since then deeper understanding has been achieved.\cite{weinberg, harvey} 

Topological protection is one of the most fascinating properties of quantum anomalies, and remarkably this protection is {\it independent}  of interaction strength.  't Hooft first realized and applied these properties  to confinement physics of quantum chromodynamics (QCD), so-called 't Hooft mathcing, and constrained candidates of low energy degrees of freedom including the Goldstone bosons from the chiral symmetry breaking.\cite{matching}
Such non-perturbative nature has been extensively applied to high energy physics, for example, the standard model, the Skyrme model of hadrons, and black hole physics.  \cite{weinberg,harvey,matching}
In condensed matter systems,        
 it is also applied to several topological phases. \cite{wen, ludwig, dirac, balents,frame, spt1, spt2,  spt3,spt4,spt5,spt6,spt7,spt8,spt9,masaki,wiegmann} For example, the presence of edge states in quantum Hall states and violation of  the chiral current conservation in Weyl semimetals are examples of the realizations of the chiral $U(1)$ anomaly.
Massless excitation in either edge or bulk is protected by anomalies' topological nature.
 
In this paper, we consider another realization of quantum anomalies in condensed matter systems, non-abelian anomalies in quantum phase transitions between competing orders and investigate consequences of their presence.  We first show that a class of competing order theories has anomalies. Their presence becomes criteria to characterize competing order theories. With the criteria,  we find that the competing order theory with the $CP(1)$ deconfined transition in three spatial dimensions cannot describe the same universality class of the NLSM with WZW in sharp contrast to the case of two spatial dimensions where the two models are proposed to describe the same universality class.  
Furthermore, by using the 't Hooft anomaly matching condition, we also find competing order physics with anomalies  {\it must} contain massless excitation in sharp contrast to the case of the conventional Landau-Ginzburg-Wilson theory.   
We provide possible candidate theories of the quantum phase transitions with anomalies.  
\\

{\textbf{\large Theories of Competing Orders }} 

Various order parameters appear in strongly correlated systems, and intriguing interplay between order parameters has been reported.\cite{comp0, lee, hirsch, chubukov, gegenwart, comp1,comp2,comp3} In this section, we introduce two types of competing order theories, which describes fundamentally different competing order mechanisms, and set up our notation for later discussion. It is worthwhile to mention that we only focus on competing order theories in non-metallic systems in this paper.  

\begin{figure}
\includegraphics[width=6.5in]{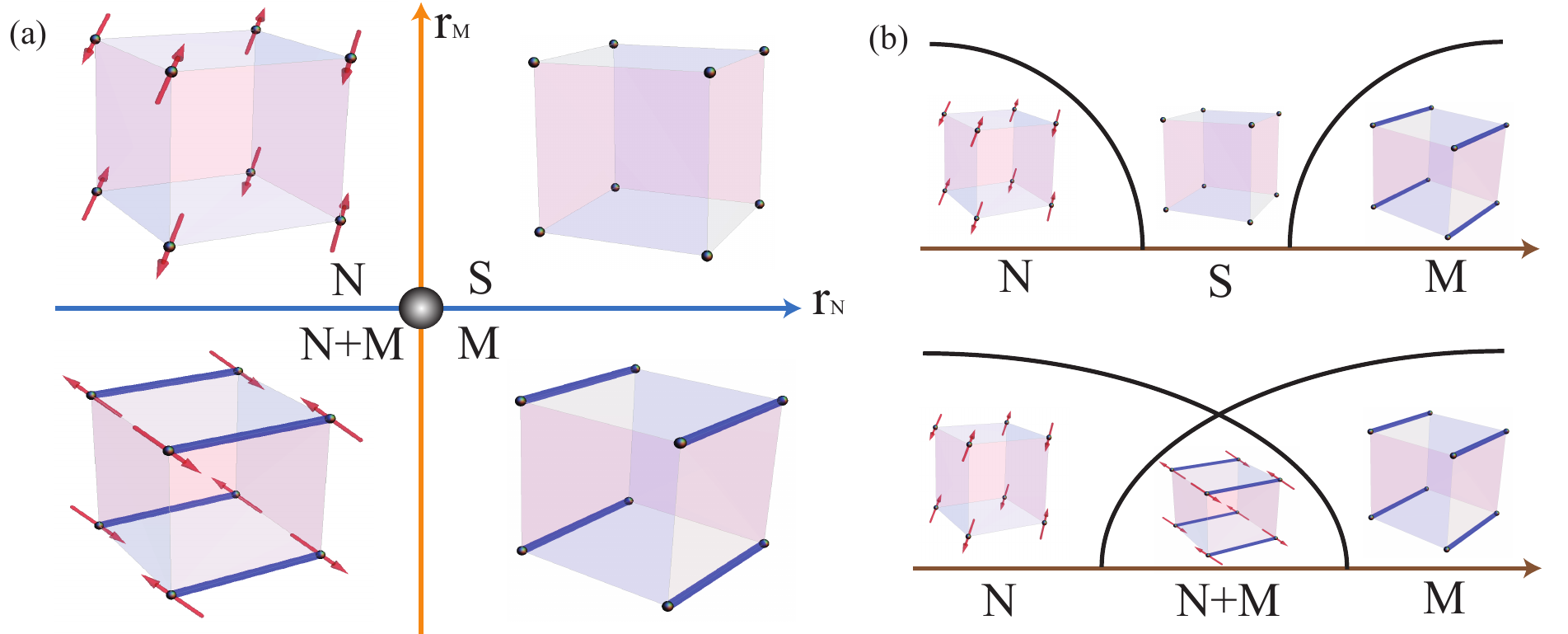}
\caption{ {(a)}  Schematic phase diagram of two order parameters in LGW. 
The horizontal and vertical axes are tuning parameters of the order parameters. 
Four phases are determined by the signs of  ($r_N$, $r_M$) :
$S$ phase  $(\langle \varphi_N \rangle=\langle \varphi_M \rangle=0)$, $N$ phase $(\langle \varphi_N \rangle\neq 0, \langle \varphi_M \rangle=0)$, $M$ phase $(\langle \varphi_N \rangle=0, \langle \varphi_M \rangle\neq 0)$, and $N+M$ phase $(\langle \varphi_N \rangle\neq 0, \langle \varphi_M \rangle \neq 0)$.
The multi-critical point ($0,0$) is well described by a theory with an enlarged symmetry group ($O(N+M)$).  
For illustration, we choose $N=3$ and $M=3$ (for example, magnetism and valence bond solid) in three spatial dimensions.
(b) Two generic phase diagrams with one parameter. The vertical axis is for magnitude of order parameters.  
The upper (lower) one is realized with the condition $r_N + r_M >0 \,(<0)$, and the fine-tuned condition ($r_N+r_M=0$) gives a second order phase transition between the two symmetry broken phases.
} \label{PD}
\end{figure}

The phenomenological $\varphi^4$ theory, so-called the Landau-Ginzburg-Wilson (LGW) theory, is one of the simplest ways to describe competing orders. 
Each order parameter ($\varphi_{N}$, or $\varphi_M$) is a representation of corresponding symmetry group (say,  vector representations of $O(N)$ or $O(M)$). 
The minimal Landau functional is 
\begin{eqnarray}
\mathcal{F}_{L}= r_{N} \varphi_N^2 +u_{N} \varphi_N^4 + r_{M} \varphi_M^2 +  u_M \varphi_M^4 + s \varphi_N^2 \varphi_M^2, \nonumber
\end{eqnarray}
omitting fluctuation terms and higher order terms. 
Four different phases are basically described by the signs of the tuning parameters ($r_N$, $r_M$) around the multi-critical point $(0,0)$ as shown in FIG. \ref{PD}. 
Note that low energy excitation of  the LGW theory's symmetric ground state (S) is massive which can be easily shown by restoring fluctuation terms.  
One useful way to understand the phase diagram is to promote the symmetry   $O(N) \times O(M) $ to $O(N+M)$  symmetry by introducing a `super-spin' ($\varphi_N, \varphi_M$) and focus on the multi-critical point. Then, the four phases are accessed by introducing anisotropy operators. 
It is well-known based on symmetry that the $\varphi^4$ theory is equivalently described by the non-linear sigma model (NLSM)\cite{subir},
\begin{eqnarray}
\mathcal{S}_0 = \int d^D x \frac{1}{2 g^2} (\partial \phi_i)^2, \quad  \sum_{i=1}^{N+M} \phi_i^2=1,  \nonumber
\end{eqnarray} 
$g$ characterizes strength of fluctuations, and the four phases are again accessed by anisotropy operators near the critical coupling constant ($g= g_c$).

Another type of competing orders theories can be obtained by incorporating topological terms to NLSMs. It is because  NLSMs can easily include topological natures of the compact ground state manifold ($S^{N+M-1}$).     
There are two types of topological terms in NLSMs : 
NLSMs with the Wess-Zumino-Witten (WZW) term and NLSMs with the $\Theta$ term. The former is realized in $D = N+M-2$, which is our main focus in this paper, and the latter is realized in $D=N+M-1$, which will be discussed in future work.  
 
The $O(D+2)$ NLSMs with the WZW terms in $D$ space-time  dimensions are  
\begin{eqnarray}
\mathcal{S}=\int_{X_D} \frac{(\partial \phi)^2}{2 g^2}+ k \,S_{WZW}, \quad \sum_{i=1}^{D+2}\phi_i^2 =1, 
\end{eqnarray}
with the spacetime manifold $X_D$.
The ground state manifold is $S^{D+1}$. 
The WZW term  is
\begin{eqnarray}
S_{WZW} &=& i c_{D+1} \int_{X_{D+1}} \frac{\varepsilon_{i_1 \cdots i_{D+2}} \varepsilon^{\mu_1 \cdots \mu_{D+1}} }{(D+1)!}\phi^{i_1} \partial_{\mu_{1}} \phi^{i_2}  \cdots,  \nonumber 
\end{eqnarray}
and the presence of the anti-symmetric tensor is related to the winding number of the ground state manifold, $\pi_{D+1}(S^{D+1})=\mathbb{Z}$.
We mainly focus on $D=2,4$ though its generalization is straightforward. 
The boundary of the manifold $X_{D+1}$ is $\partial X_{D+1} = X_D$. The numerical constants are $c_{D+1} = \frac{2\pi }{ {\rm Area}(S_{D+1})}$ and $ {\rm Area}(S_{D+1}) = \frac{2 \pi^{(D+2)/2}}{\Gamma((D+2)/2)}$.
 ($c_3 = \frac{1}{\pi}$ and $c_5 = \frac{2}{\pi^2}$). 
With a non-zero $k$, this class of competing order theories have been suggested to describe exotic competing order physics, for example, competition between Neel and valence bond solid orders associated with deconfined quantum criticality.\cite{senthil, ashvin}

The engineering dimension of the coupling constant is $[g^2] = 2-D$. In $D>2$, the coupling constant is irrelevant indicating the model is well-suited to describe a weak coupling fixed point, a symmetry broken phase. Therefore, to access symmetric ground states, one need to consider strong coupling limits where a perturbative calculation is not reliable. Below, we show that non-perturbative nature of quantum anomalies allows us to investigate symmetric phases even in strong coupling limits.

In $D=2$, the coupling constant is marginal at the tree level, and Witten showed the WZW-NLSM is mapped to a massless fermion model by non-abelian bosonization\cite{witten}. 
This clearly shows that the WZW-NLSM in $2D$ describes a different universality class from the LGW $\varphi^4$ theory  whose ground state has energy gap as shown by Mermin and Wagner.\cite{subir} 
Thus, it is clear that the WZW term plays a crucial role to modify the ground state in a symmetric phase. 
Soon after, it was understood that the spin $\frac{1}{2}$ chain is described by the $ O(4)$ WZW-NLSM, and the onset of the valence-bond-solid order is understood by a marginal perturbation, which is interpreted as competition between spin and valence-bond-solid order. \citep{senthil} 
Based on the $2D$ results, the WZW-NLSMs in higher dimensions are suggested to be in a different universality class from the LGW theory applying to various competing order physics. \cite{tanaka,senthil, senthil2, ashvin, lee1, lee2, moon2,herbut,sudip,si,fu,fradkin1,fradkin2}

It is, however, significantly more difficult to analyze the models in higher dimensions than 2D partially due to a lack of a local conformal symmetry.  Furthermore, a continuous symmetry can be spontaneously broken in higher dimensions, so a symmetry-broken state becomes another stable fixed point, which makes  renormalization group flows complicated. 
The powerful theoretical tools such as $\epsilon= D-2$ or large $N$ methods are not applicable since changing space-time dimensions and an order parameter manifold are prohibited by the presence of the WZW terms.    
Therefore, it is quintessential to find ways to understand properties of symmetric phases in the WZW-NLSMs. Below, we find one concrete way to investigate symmetric phases of the WZW-NLSM type competing order physics at least in even-spacetime dimensions. 
\\

{\textbf{\large Anomalies in WZW-NLSM}}

We employ the standard strategy to investigate quantum anomalies: to promote a global symmetry to a local one (gauging) and search for inconsistency. \citep{weinberg,harvey} The promotion is done by introducing a gauge potential and associate minimal coupling.
Without the WZW term, the minimal coupling is enough to gauge the symmetry and there is no ambiguity. 
But, gauging the WZW term is subtle since it is expressed in $X_{D+1}$.
One criterion of a proper gauging procedure is that gauging does not change space-time dimensions of dynamics. For example, equations of motions should be well-defined in the original dimensions ($X_D$) even after gauging since infinitesimally weak gauge coupling is conceivable. \cite{hull}  

For notation convenience,  we use differential forms following Nakahara's\cite{nakahara}  (also, see Supplementary Materials).
In $2D$, the WZW term is 
\begin{eqnarray}
S_{WZW}=  i c_3 \int_{X_3} \omega_3, \quad \omega_3= \frac{ \varepsilon_{ijkl}}{3!}  \,\phi^i d \phi^j \wedge \, d\phi^k \wedge \, d\phi^l. \nonumber
\end{eqnarray}
Note that even though the WZW term is expressed in $X_{3}$, the equation of motion is well-defined in $2D$,  
\begin{eqnarray}
-\frac{1}{g^2} \partial^2 \phi^i + \frac{i k}{2\pi}    \varepsilon^{\mu \nu} \varepsilon_{ijkl} \phi^j  \partial_{\mu}\phi^k \partial_{\nu}\phi^l =0.
\end{eqnarray}
This is because $\omega$ is the highest form ($d \omega =0$) in $X_3$, and the closeness ($d \omega =0$) is tied to a two dimensional equation of motion from Poincare's lemma. 
From now on, we focus on $SO(D+2)$ symmetry instead of $O(D+2)$ since the 't Hooft anomaly matching only cares continuous groups. 
The variation of the order parameter in the $SO(4)$  vector representation is $\delta \phi^i =\epsilon^{a}[t_a]^i_j \phi^j $
with the Lie algebra  $[t_{a}]_{ij}=-[t_{a}]_{ji}$ of $SO(4)$, ($a=1,\cdots,6$). 
Local dependence of  $\epsilon^a(x)$ determines whether the transformation is global or local.

We introduce  a gauge potential $A = A_{\mu}^a t_a dx^{\mu}$ to gauge the $SO(D+2)$ symmetry whose field strength is $F^a =d A^a + \frac{1}{2} f^a_{bc} A^b \wedge A^c 
$.
The covarinat derivative is 
\begin{eqnarray}
D\phi^i \equiv  (d+A)\phi^i = D_{\mu}\phi^i dx^{\mu}= (\partial_{\mu} \delta_{ij} + A_{\mu}^a [t_a]_{ij})\phi^j dx^{\mu}. \nonumber
\end{eqnarray}
Then, the minimally coupled WZW term in $2D$ is 
\begin{eqnarray}
\widehat{S}_{WZW} = i c_3 \int_{X_3} \widehat{\omega}_3, \quad \widehat{\omega}_3=\frac{ \varepsilon_{ijkl}}{3!}   \,\phi^i D \phi^j \wedge \, D\phi^k \wedge \, D\phi^l. \nonumber 
\end{eqnarray}
The ($\,\widehat{}\,$) notation is for the minimal coupling with a covariant derivative ($\Omega(\partial) \rightarrow \widehat{\Omega} = \Omega(D) $).
Gauge transformations ($\delta_g$) of the potential and the covariant derivative are
$\delta_g A^a = -d\epsilon^a + f^a_{bc}A^b \epsilon^c$ and $\delta_g (D\phi^i) = \epsilon^a(x) [t_a]_{ij} (D\phi^j)$. 
By construction, the minimally coupled WZW term is gauge-invariant.

Its equation of motion from the minimal coupling, however, is not two dimensional. 
This is easily shown by an exterior derivative of the minimally coupled WZW form and we find 
\begin{eqnarray}
d \widehat{\omega}_3 
&=&-  d \Big[ F^a \wedge \widehat{v_a}+ d_{ab} \,(Q_3)^{ab} \Big] \neq 0. \nonumber 
\end{eqnarray}
An one-form $v_a = \frac{1}{4} \varepsilon_{ijkl} [t_a]_{ij} \phi^k d \phi^l$ and the three dimensional Chern-Simon (CS) term,
$(Q_3)^{ab}= A^a \wedge(d A^b + \frac{ f^{b}_{cd} A^c \wedge A^d }{3})$ are introduced. 
The anomaly coefficient is 
\begin{eqnarray}
d_{ab} = \frac{ \varepsilon_{a_1 a_2 {b_1}  b_2}}{ 4}. \nonumber
\end{eqnarray}
Thus, the simple minimal coupling is not enough to make the WZW term gauge-invariant and the equation of  motion two-dimensional.  

To make the equation of motion well-defined in $2D$, one can rearrange and find the closed form. The wedge product ($\wedge$) is implicit hereafter.
\begin{eqnarray}
d  \Big[\widehat{\omega}_3 +F^a \widehat{v_a}+ d_{ab} \,(Q_3)^{ab} \Big]=0. \nonumber 
\end{eqnarray} 
Thus, the gauged total action with a two dimensional equation of motion should be 
\begin{eqnarray}
S_{tot} = \widehat{S}_0+\widehat{S}_{WZW} + (i \,c_3) k \int_{X_3} \Big( F^a \widehat{v_a} + d_{ab} \, Q_3^{ab} \Big).
\nonumber
\end{eqnarray}
Note that the last two terms contain gauge potential and field strength, so they vanish without gauge fields.

The gauged action, however, has significant inconsistency under gauge transformations
\begin{eqnarray}
&&\delta_g S_{tot}= - (i \,c_3) k \int_{X_2}  d_{ab} \, d\epsilon^a A^b,
\end{eqnarray}
whose origin is the presence of the Chern-Simon (CS) term. Therefore, it is {\it impossible} to gauge the $SO(4)$ symmetry in the WZW-NLSM, which indicates the symmetry is anomalous. 
Note that similar inconsistency appears at the boundaries of the $U(1)$ CS theory in quantum hall systems.\cite{wen} 
 
In $4D$ with the $SO(6)$ model, the similar procedure is applied with little more tedious calculation. 
The five dimensional volume form is 
\begin{eqnarray}
\omega= \frac{\varepsilon_{ijklmn}}{5!}\, \phi^i\, d\phi^j \,d\phi^k\, d\phi^l\, d\phi^m\, d\phi^n, \nonumber 
\end{eqnarray}
and the exterior derivative of the minimally coupled WZW term is
 \begin{eqnarray}
d \Big( \widehat{\omega}  + F^a \widehat{v_a}+F^a F^b v_{(ab)} + d_{abc} Q_5^{abc} \Big) =0\nonumber
\end{eqnarray}
with the anomaly coefficient 
\begin{eqnarray}
d_{abc} =  \frac{\varepsilon_{a_1 a_2 b_1 b_2 c_1 c_2}}{4!}. \nonumber
\end{eqnarray}The five dimensional CS term is 
$(Q_5)^{abc}= A^a d A^b dA^c+ \frac{3}{4}f^c_{ef}A^e A^f A^a dA^b +
\frac{3}{20} A^a f_{de}^b A^d A^e f_{gh}^c A^g A^h)$, and the three form is $v_a = \frac{1}{32} \varepsilon_{ijklmn} (t_a)_{ij} \phi^k d \phi^l d \phi^l d\phi^m$. The symmetrized interior derivative is introduced, 
 $ i_{(a} v_{b)}=d v_{(ab)}$.
It is straightforward to show the $SO(6)$ symmetry is anomalous.

Note that the anomaly coefficient is only non-zero when all the indices are used up. 
This indicates that gauging the full $SO(4)$ in $2D$ and $SO(6)$ in $4D$ is crucial to find anomalies. 
Gauging subgroups such as $SO(3)$ does not use up all the indices, so the coefficient automatically vanishes. In other words, gauging subgroups is always well-defined, and the anomaly structure only appears when the full symmetry is gauged.

 
Three remarks follows. First, our anomaly coefficient calculation in WZW-NLSMs is {\it independent} of the coupling constant strength ($g^2$). This is consistent with non-perturbative nature of quantum anomalies : in the extreme limit with a very large (bare) coupling constant, one can imagine  a symmetric ground state. Still, the anomaly structure must be there, so there must be massless excitation to reproduce the anomaly structure since order parameters are completely energy-gapped in symmetric phases.    
The presence of massless excitation can be more rigorously shown by investigating singularity structure of currents correlation functions\cite{coleman, frishman,coleman2} (see also Supplementary Materials).    
Second, the presence of massless excitation in $2D$, which is consistent with the Witten's bosonization results, and $4D$ with anomalies indicates their ground states are qualitatively different form LGW theory's. Thus, our anomaly calculation shows the WZW-NLSMs describes massless symmetric phases qualitatively different from the LGW $\varphi^4$ theory and provides non-perturbative criteria to distinguish quantum criticalities. 
Third, our calculation can also be applied to relations between topological boundary and bulk phases with non-abelian symmetries. The gauge-invariance and equation of motion properties are intrinsically connected through the presence of the $D+1$ dimensional Chern-Simon terms. Their presence appears in a certain class of  D+1 dimensional topologically ordered phases. Therefore, our work explicitly shows that the D+1 dimensional (non-abelian) topological phases, related with the non-abelian Chern-Simon theories, can have $D$ dimensional gapless boundaries guaranteed by the presence of quantum anomalies.   
\\
 
\textbf{\large Anomaly matching and Minimal model} 

Quantum anomalies guarantees the presence of massless excitation in symmetric phases but it does not pin down the symmetric ground state completely. 
But, at least, it is obvious that the conventional strong coupling analysis as in the LGW theory is not applicable to the competing order theories with the WZW term. 

A priori, it is not even clear how many phase transitions are and whether they are first or second orders in strong coupling limits. It is because all ignored higher order terms become important in strong coupling limits. Thus symmetric ground states of the WZW-NLSMs are not uniquely determined without further microscopic information. But, no matter what happens, the presence of massless excitation is guaranteed by quantum anomalies.   

We first consider one specific model which reproduces a given quantum anomaly structure which helps us to understand qualitative differences better. Then, we use 't Hooft matching condition to investigate candidates of symmetric phases.    
The minimal model to describe a symmetric phase with quantum anomaly can be obtained from the symmetry breaking pattern in the symmetry-broken phases, $SO(6)/SO(5) \sim S^5$ in $4D$. 
It is well-known in literature that the two color $(N_c=2)$ QCD with the two flavor $(N_f=2)$ enjoys the enlarged $SU(4)$ flavor symmetry instead of $SU(2)_L \times SU(2)_R$. \cite{qcd1,qcd2, hill} 
The symmetry is spontaneously broken to $Sp(4)$ by the chiral condensate. Since $SU(4)$ is isomorphic to $SO(6)$ and $Sp(4)$ to $SO(5)$ in terms of the Lie algebra, the symmetry breaking pattern is exactly  $SO(6)/SO(5)$ and the dynamics of the Goldstone boson is described by a $SO(6)$ NLSM (see also Supplementary Materials). 
In 2D, it is well known that  spin $1/2$ chains realize the $SO(4)$ WZW-NLSM.

Inspired by the hints, we construct fermion models coupled to the $SO(D+2)$ bosons. 
We introduce a complex spinor, $\Psi$, which couples to the order parameters,
\begin{eqnarray}
S_f &=& \int \Psi^{\dagger} (\partial_{\tau} -i  \gamma^{s}\partial_{s})\Psi + \lambda \phi^i \Psi^{\dagger} \Gamma_i \Psi.
\end{eqnarray}
$s$ is for spatial dimensions, and $i$ is for order parameters.
$(\gamma^s, \Gamma_i)$ matrices satisfy the Clifford algebra.  $s=1\, (1,2,3)$, and $i=1,\cdots, 4$ $(1\cdots6)$ in 2D (4D). 
By the Yukawa coupling, the fermions become massive in the symmetry broken phases. 
The minimum numbers of spinor components for the Clifford algebra are four in $2D$ and sixteen in $4D$ as in the spin chain and $N_c=2$ QCD.

The WZW term can be easily reproduced in the symmetry broken phases by integrating out the fermions,
\begin{eqnarray}
\Gamma^{eff}= -\log \mathcal{Z}, \quad  \mathcal{Z}[\phi] = \int_{\psi, \psi^{\dagger}} \exp{(-S_f)} = {\rm Det} \mathcal{D} \nonumber
\end{eqnarray}
with $\mathcal{D}=\partial_{\tau} -i \gamma^{s}\partial_{s}+\lambda \phi^i \Gamma_i$. 
Note that matching the level of the WZW term automatically satisfies the anomaly matching condition, and we indeed find the same level by integrating out the fermions. In Supplementary Materials, standard field theoretic consideration with group structures\cite{weinberg, harvey} is presented to be self-contained which is useful in symmetric phases. 
It is also straightforward to show further gradient expansion in the symmetry broken phase of the minimal model reproduces the WZW-NLSM.

%

The minimal model in $4D$ is naturally constructed by fermions and bosons, 
\begin{eqnarray}
S_{min} &=& \int \Psi^{\dagger} (\partial_{\tau} -i \gamma^{r}\partial_{r})\Psi + \lambda \phi^i \Psi^{\dagger} \Gamma_i \Psi \nonumber \\
&+&\int \frac{1}{2} (\partial \phi^i)^2 + \frac{r}{2}  (\phi^i)^2 +\frac{u}{4!}  ((\phi^i)^2)^2.
\end{eqnarray} 
$\Psi$ is in $SO(6)$ spinor representation (sixteen complex fields) and $r$ is a tuning parameter. 
In $2D$, a similar model can be constructed  which is well understood, so we focus on $4D$ from now on.
By tuning $r$, one can access the symmetry broken phase ($r < r_c$) and the symmetric phase ($r > r_c$) where $r_c$ is a critical value and its numerical value depends on a renormalization group scheme. 
The minimal model is a class of the so-called Higgs-Yukawa theory, and uniqueness of the minimal model is  the symmetry structure of bosons and fermions associated with quantum anomalies.  
 
The renormalization group flow of the general Higgs-Yukawa theory is well understood\cite{srednicki}, and our minimal model has the same structure.     
The model is at the upper critical dimension $4D$, so the mean-field description is valid with logarithmic corrections.
There are three fixed points : the symmetry broken phase $N+M$, the quantum critical point $QC$, and the symmetric phase $S$ (fermionic) as illustrated in the horizontal line of Fig. 2.
Note that the symmetry broken phases in the minimal model and the LGW theory are similar, but quantum critical points and symmetric phases are fundamentally different due to the presence of fermions. 

\begin{figure}
\includegraphics[width=5.in]{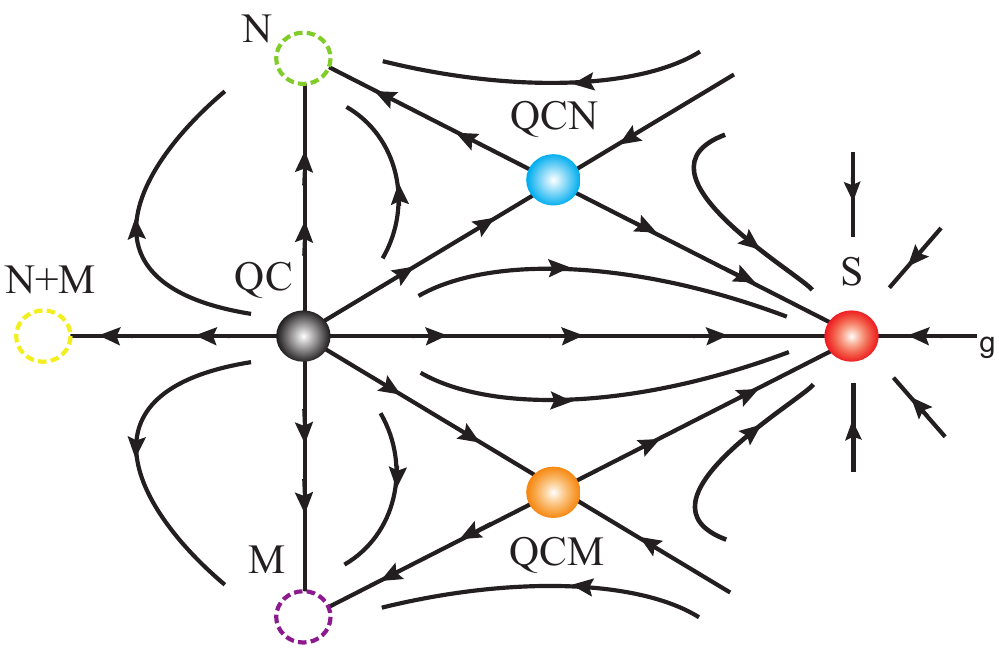}
\caption{  Schematic RG flow of the minimal model in 4D with two parameters. The horizontal (vertical) axis is to characterize fluctuation (anisotropy) strength. 
$N+M$ is $SO(6)$ symmetry broken phase characterized by $5$ Goldstone modes. 
$M$ ($N$) is $SO(M)$ ($SO(N)$) symmetry broken phase characterized by $M-1$ ($N-1$) Goldstone modes. 
$S$ is $SO(6)$ symmetric phase with massless excitation  enforced by quantum anomalies.  
} \label{white}
\end{figure}


We consider perturbations which break the full $SO(6)$ symmetry down to its subgroup $ H=SO(N) \times SO(M)$, with $N+M=6$  to connect the anomaly structure with competing orders (we treat $SO(1) \sim \mathbf{Z}_2$).
 For example, $SO(3) \times SO(3)$ symmetry in the minimal model  allows anisotropy operators 
\begin{eqnarray}
\sum_{i=1,2,3}(\phi^i)^2 - (\phi^{i+3})^2, \quad \sum_{i=1,2,3}(\Psi^{\dagger} \Gamma^i \Psi)^2 -(\Psi^{\dagger} \Gamma^{i+3} \Psi)^2. \nonumber
\end{eqnarray}
Again, the minimal model is at the upper critical dimension, so one can easily read off scaling dimensions of the operators at each fixed point. 
Near $QC$, the former operator is relevant because it reduces the number of massless modes but it is irrelevant at $S$ since the order parameters are gapped. 
The four point fermion interaction is irrelevant in all three fixed points. 
Schematic RG flow of the minimal model with the anisotropy parameter is straightforwardly obtaind as  in  Fig. 2. 
Different symmetry breaking patterns with different $N, M$ give similar RG flows.  
Note that the RG flow structure of the minimal model is similar to one of the LGW theory, but crucial distinction between the minimal model and the LGW theory is the presence of massless fermions in S enforced by quantum anomalies. 
If the fermions are identified as electrons, the symmetric phase is nothing but Weyl or Dirac semi-metals.\cite{dirac}

Before closing this section, we emphasize that the reason we consider the minimal model is to provide a concrete example of the symmetric phases. But, our anomaly calculation is powerful enough to be applied to more generic cases even with strongly correlated ground states. We discuss such more generic cases below.
\\

\textbf{\large Non-minimal models}

%

Let us consider non-minimal models which have the same anomaly structure.  
A priori, all conformal field theories (CFTs) with the same anomaly coefficient are candidate theories of S.
One straightforward way to construct non-minimal models is to consider different representations of $SU(4)$ fermions instead of a single $SU(4)$ fundamental representation of the minimal model. 
Detailed discussion about other representations is presented in Supplementary Materials. 
Notice that the RG flow structure of the models with different representations is basically the same as the minimal model's.

If fermions are not electrons but fractionalized particles such as spinons, then the symmetric fixed point can be identified as a spin-liquid phase.  
If spinons are weakly coupled to gauge field such as $U(1)$ or $Z_2$, then all the properties of the electronic minimal model is inherited and the symmetry broken phases contain remaining gauge structure, so-called $*$ phases. They do not describe conventional symmetry broken phases. 
Therefore, if the symmetric fixed point with spinons are adjacent to conventional (confined) symmetry broken phases, the gauge structure must be non-abelian. 
For example, the well-known Banks-Zak fixed point \cite{banks} with spinons could be a candidate of the symmetric fixed point S. Then, condensing the order parameter endows spinon energy gap, and the remaining gauge field becomes confined naturally.

\begin{figure}
\includegraphics[width=5.5in]{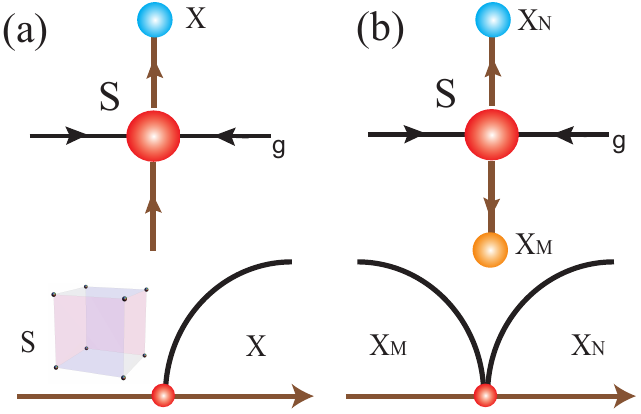}
\caption{Marginal and relevant operators with $H = SO(N) \times SO(M) $ symmetry. 
(a) RG flow with a marginal operator around the symmetric phase (top). The horizontal (vertical) axis is to characterize fluctuation (anisotropy) strength.  
Phase diagram with a marginal operator (bottom). Sign of the tuning parameter determines relevance (or irrelevance). Thus, one side of phase diagram is a $SO(6)$ symmetric CFT,  and the other side ($X$) is either a $H$ symmetric CFT or a $H$ symmetry broken phase.
(b) RG flow with a relevant operator around the symmetric phase. The relevant operator drives the RG flow from $S$ to $X_M$ and $X_N$. They are either $H$ symmetric CFTs or $H$ symmetry broken phases. 
Once $X_M$ and $X_N$ break different symmetries, the symmetric fixed point ($S$) may describe a ``deconfined'' quantum criticality. 
The curved (black) lines in the bottom figures are for energy scales associated with symmetry breaking (e.g. order parameter scale).
} \label{marginal}
\end{figure}

In principle, two different RG flow structures near S are possible if the symmetric fixed point S is described by  strongly coupled CFTs with the same anomalies.
In contrast to the minimal model's RG flow, anisotropy operators could be relevant or marginally relevant. 
If marginally relevant, then the symmetric phase becomes stable with one definite sign of the coupling constant, but the opposite sign makes the symmetric phase unstable. Thus, one side of phase diagram is described by a $SO(6)$ symmetric CFT, and the other side ($X$) is described by  either a $H$ symmetric CFT or a $H$ symmetry broken phase as shown in FIG. \ref{marginal}(a).
If relevant, then the symmetric phase becomes unstable with both signs of the coupling constant.
Again, the final states can be either $H$-symmetric CFTs or $H$-broken phases as shown in FIG. \ref{marginal}(b).
Once the final two states break different symmetries, then the symmetric fixed point ($S$) connects two broken phases directly, which describes deconfined quantum criticality.\cite{deconfined}

The above discussion gives the two necessary conditions to realize deconfined quantum criticality in $SO(6)$ WZW-NLSM in $4D$ : anomalies and relevant symmetry breaking operators. 
These conditions provide non-perturbative criteria to characterize universality class of quantum phase transtions between competing orders. 
For example, the universality class of the $SO(6)$ WZW-NLSM cannot be the same as one of the non-compact $CP(1)$ model in $4D$ due to the absence of anomalies in the latter model. Notice that in $3D$ the $CP(1)$ model and the $SO(5)$ WZW-NLSM are proposed to describe the same universality class but our anomaly criteria do not applied to odd space-time dimensions.  \\

\textbf{\large Discussion and  Conclusion}

In experiments, direct measurement of the non-abelian anomalies associated with competing orders is even more difficult than one of the chiral $U(1)$ anomaly in Weyl semi-metals because we do not know how to couple the non-abelian current directly in experiments.\cite{hosur, burkov} 
Yet, there are traits associated with the anomalies.

Protection of massless excitation is one of the most significant characteristics of the presence of quantum anomalies. Their numbers are, however, not universal.    
For example, in the minimal model in $4D$, the numbers of massless excitation in Goldstone phase, quantum critical point, and symmetric phase are $5$ (bosons), $21 (=6+16)$ (bosons +fermions), and $16$ (fermions). 
In non-minimal models, the symmetry broken phase has the same number, but critical point and symmetric phase have different numbers of massless excitation. Clearly, this is different from LGW theory's where all massless excitation has definite numbers at each fixed point. The different numbers of massless excitations may contribute to transport differently, which is in principle measurable.

In the minimal model, characteristics of anomaly becomes more evident.
First of all, semi-metallic behaviors appear in a symmetry restored phase or high temperature regime (but lower than cut-off scale, say band width) if fermions are electrons.    
Massless electrons in a symmetric phase (or quantum critical regime) governs low energy physics, so electrical and thermal currents are carried by electrons with the linear spectrum.  Naturally, the Wiedemann-Franz law holds especially in non-hydrodyanmic limits.  
By lowering temperature, the $SU(4)$ symmetry can be broken and the electrons become gapped.  
Thus, electrically, insulating behaviors (energy gap) appear, and the symmetry breaking transition is concomitant with the transition between semi-metal and insulator.
On the other hand, Goldstone modes from spontaneous symmetry breaking carry thermal currents even though electrons are gapped.
Since Goldstone modes and massless electrons have same dispersion relation, thermal transport in symmetry broken phases is qualitatively similar to the one in symmetric phases.    
Thus, near the semi-metal and insulator transition, electric and thermal currents behave differently, and the Wiedemann-Franz law would be violated.  


One experimentally realizable system of the $SU(4)$ anomaly is pyrochlore systems with all-in-all-out magnetic order parameter\cite{pyrochlore} in addition to $N_c=N_f=2$ QCD . 
A class of pyrochlore structure is described by a quadratic band touching model \cite{qbt} and the onset of all-in-all-out order parameter \cite{aiao} induces eight Weyl points (16 fermions), the minimal necessary number to realize the $SO(6) \sim SU(4)$ anomaly. 
We note that evidence for all-in all-out ordering and violation of the Wiedemann-Franz law in spin-orbit coupled pyrochlore structures has been reported in literature \cite{mandrus,satoru} though precise connection with anomalies need further investigation.  

In this paper, we investigate non-abelian anomalies in quantum phase transitions with competing orders. We show that the WZW-NLSMs in $2D$ and $4D$ have quantum anomalies by calculating the anomaly coefficients.    
Non-perturbative nature of the anomalies allows us to investigate not only a symmetry broken phase in weak coupling limit but also a symmetric phase in strong coupling limit even though the presence of the WZW term prohibits conventional $\epsilon=D-2$ and large $N$ expansion methods.  Applying the 't Hooft matching condition, it is shown that the universality class of the models is qualitatively different from the conventional  $\varphi^4$ theory's. In sharp contrast to the $\varphi^4$ theory, symmetric ground states of WZW-NLSMs contain massless excitation though their numbers are not uniquely determined. Thus, we construct the minimal model and investigate its properties under anisotropy perturbation. Then, we extend the  model to more general ones and discuss implication of the anomalies in competing order physics.  Further research on anomalies, for example, parity anomaly in odd space-time dimensions and mixed anomalies with gravity (thermal effects) in connection with topological phases (NLSM with the theta term) are desirable.

{\it Acknowledgements:}
It is great pleasure to discuss with M. Goykhman, L. Kadanoff, D. Kutasov, S. Lee, S. S. Lee, M. Levin, E. Martinec, M. Roberts, D. T. Son, and P. Wiegmann. 
The author is of particular grateful to D. T. Son for introducing 't Hooft anomaly and $N_c=2$ QCD and to S. Lee for discussion about field theory and differential  geometry.  
This work is supported by the Kadanoff Center Fellowship and the KAIST start-up funding.




\appendix

\section{Massless excitation with anomalies}
It is well understood that massless excitation is guaranteed by continuous symmetry anomalies.\cite{coleman, frishman, coleman2}
The presence of continuous group's anomalies enforces singularities of analytical structures of currents correlation functions. To be self-contained, we introduce the proof with slight modification following the notation in Coleman and Grossman\cite{coleman2}.

In $4D$, the anomalous Ward identity is in three currents correlation function, 
\begin{eqnarray}
&&\Gamma_{\mu \nu \lambda}(q_1, q_2, q_3)  \delta^{(4)}(q_1+q_2+q_3) = \nonumber \\
&&\int \prod_i d^4 x_i e^{i q_i x_i} T<0|J_{\mu}(x_1) J_{\nu}(x_2) J_{\lambda}(x_3)|0>, \nonumber
\end{eqnarray} 
and the current conservation gives
\begin{eqnarray}
q_3^{\lambda} \Gamma_{\mu \nu \lambda}(q_1,q_2,q_3) = \mathcal{A} \epsilon_{\mu \nu \alpha \beta} q_1^{\alpha} q_2^{\beta}. \label{anomaly eqn}
\end{eqnarray}  
All non-abelian Lie algebra indices are absorbed into the anomaly coefficient $\mathcal{A}$.

The correlation function is symmetric under simlutaneous permutations of $(q_1, q_2, q_3)$ and $(\mu, \nu, \lambda)$. 
Now let us investigate analytic structure of the correlation function. Due to permutation and covariance, the structure must be in the form
\begin{eqnarray}
\Gamma_{\mu \nu \lambda} &=& F(q_i^2) \Big[ \epsilon_{\mu \nu \alpha \beta} q_1^{\alpha} q_2^{\beta} q_{3\lambda} + \epsilon_{ \nu \lambda \alpha \beta} q_2^{\alpha} q_3^{\beta} q_{1 \mu} +\epsilon_{\mu \nu \alpha \beta} q_3^{\alpha} q_1^{\beta} q_{2 \nu} \Big] \nonumber 
\end{eqnarray}
We omit possible tensors which cannot contribute to the anomalies. 
Note that the momentums are off-shell, so one can access all available regions and we focus on the region 
\begin{eqnarray}
q_1^2 = q_2^2 = q_3^2 = -Q^2. \nonumber
\end{eqnarray}

The correlation function contracted with $q_{3 \lambda}$ gives
\begin{eqnarray}
q_3^{\lambda} \Gamma_{\mu \nu \lambda}(q_1,q_2,q_3) = - F(Q^2)Q^2 \epsilon_{\mu \nu \alpha \beta} q_1^{\alpha} q_2^{\beta}.
\end{eqnarray}
Then, the anomaly equation (A.1) gives 
\begin{eqnarray}
F(Q^2) = -\frac{\mathcal{A}}{Q^2}. \nonumber
\end{eqnarray}
The pole structure at zero mass nicely show the presence of massless excitation (also see \cite{frishman} for dispersion analysis). 
The singularity even further enforces that $UV$ and $IR$ information needs to be matched.

In the paper by Coleman and Grossman, they add more conditions such as non-singularties from vertex corrections, and they conclude the helicity of massless degrees of freedom is $\pm \frac{1}{2}$, which indicates the symmetric phase is massless fermions as in our minimal model. The authors argue that the assumptions are not that strong, so it would be very interesting the conditions are proved / disproved in future research.  

The above discussion only relies on the anomaly properties and nothing more, thus it is applied to everywhere in phase diagrams. 
But, it is only applied to anomalies of continuous symmetries since the current conservation plays a crucial role. For the discrete gauge group, which is especially important in SPT physics, the presence of anomalies does not guarantee massless excitation.\cite{spt1,spt2,spt3,spt4}

We note that in $2D$, the minimal symmetry for spin $1/2$ chains to be massless is $SU(2) \times Z_2$ corresponding $SU(2)\times Z_2$ \cite{masaki} which is smaller than $SO(4) \sim SU(2) \times SU(2)$, and it is manifest some subgroups of the continuous group is enough on lattice systems, and it would be interesting to find criteria to determine the subgroups in higher dimensions.

\section{ $N_c=2$ QCD theory}

In QCD, confinement issue is subtle and it is known that chiral femrions with non-abelian gauge fields are confined in the infra-red limit. 
To connect the QCD theory to our minimal model, one needs to control one parameter to access both the chiral symmetry broken phase and the symmetric phase. 
One way to do this is to tune gluon-quarks interaction strength by adding a color doublet, flavor singlet scalar (Higgs) field $\Phi$. 
In one limit, $\Phi$ is massive and then the chiral symmetry is broken by confinement, but in the other limit, the Higgs field obtain vacuum expectation value (vev) and the gluon fields become screened. A phase transition between the two limits can be reached by dialing the Higgs vev. Due to $SU(4)$ anomalies, one now expects physical massless particles in the spectrum. They are described by colorless product of the quark field and the Higgs.  Thus, in the symmetric phase, one has massless fermions in the fundamental representation of $SU(4)$, which could be identified by our minimal model. 

Below, we provide more information on symmetry enlargement and order parameter construction in $N_c=N_f=2$ QCD suppressing gluon fluctuations.

The massless Dirac Hamiltonian of $N_c=N_f=2$ QCD in the chiral representation is,  
\begin{eqnarray}
\mathcal{H}_0= q_{L}^{\dagger} i\,\sigma^k \partial_{k}q_L - q_{R}^{\dagger} i\,\sigma^k \partial_{k}q_R.  \quad (k=1,2,3)
\end{eqnarray}
$q_{L,R}$ is a spinor with $N_f \times N_c \times 2$ components (flavor $\times$ color $\times$ Lorentz). 
The flavor indices are implicit and it is manifest that the Hamiltonian has $SU_L(N_f) \times SU_R(N_f)$.

For $N_c=2$ QCD, the symmetry is enhanced to  a bigger symmetry, $SU(2N_f)$.  
One can see this by employing the chiral particle-hole transformation for a chiral field (say, $q_R$), 
\begin{eqnarray}
q_R = (T_2 \sigma^2) (\tilde{q}^{\dagger})^T, \quad q_R^{\dagger} = -\tilde{q}^T (T_2 \sigma^2), \quad q = q_L.
\end{eqnarray}
$T_{1,2,3}$ are anti-hermitian generators for the $SU(2)$ color.
We define the corresponding Hermitian operators, $t^k= i T_k$.
In terms of a new fields $\Psi^T = (q , \tilde{q})$, the Hamiltonian is
\begin{eqnarray}
\mathcal{H}_0&=& q_{L}^{\dagger} i\,\sigma^k \partial_{k}q_L - q_{R}^{\dagger} i\,\sigma^k \partial_{k}q_R \nonumber \\
&=&  q^{\dagger} i\,\sigma^k \partial_{k} q +\tilde{q}^{\dagger}  i\,\sigma^k  \partial_{k} \tilde{q} = \Psi^{\dagger}  i \sigma^k \partial_k \Psi. \nonumber 
\end{eqnarray}
$SU(2N_f)$ symmetry is manifest with new spinors. Notice that all fermions are chiral in this representation.  

One mass term (color singlet and flavor singlet) in two different representations is 
\begin{eqnarray}
q_R^{\dagger} q_L + q_L^{\dagger} q_R &=&  -\tilde{q}^T (T_2 \sigma^2) q +q^{\dagger}(T_2 \sigma^2) (\tilde{q}^{\dagger})^T \nonumber \\
&=&  -i \Big[\tilde{q}^T (t_2 \sigma^2) q -q^{\dagger}(t_2 \sigma^2) (\tilde{q}^{\dagger})^T \Big] \nonumber\\
 &\equiv& \Psi^T \rho_2 t_2 \sigma^2 \Psi + \Psi^{\dagger} \rho_2 t_2 \sigma^2  (\Psi^{\dagger})^T
\end{eqnarray}
More mass terms can be systematically obtained by using the  Majorana spinors,
\begin{eqnarray}
\Psi = \eta_1 +i \, \eta_2, \quad \eta \equiv (\eta_1 \, \eta_2)^T
\end{eqnarray}
Then, the fermion bilinears have the form
\begin{eqnarray}
&&\Psi^{\dagger} M \Psi =  \eta^T (\frac{M-M^T}{2}\mu^0 - \frac{M+M^T}{2}\mu^2) \eta \nonumber \\
&&\Psi^T M \Psi = \eta^T M (\mu^3 + i \mu^1) \eta \nonumber \\
&&\Psi^{\dagger} M (\Psi^{\dagger})^{T} = \eta^T M (\mu^3- i \mu^1) \eta \nonumber \\
&&\eta^T M \mu^3 \eta = \frac{1}{2} (\Psi^T M \Psi + \Psi^{\dagger} M (\Psi^{\dagger})^{T} ) \nonumber \\
&&\eta^T M \mu^1 \eta = \frac{1}{2 i} (\Psi^T M \Psi - \Psi^{\dagger} M (\Psi^{\dagger})^{T} ) \nonumber 
\end{eqnarray}
with $\mu^{0,1,2,3}$ are for the Majorana space. 
Thus, the mass term in the Majorana representation is 
\begin{eqnarray}
&&q_R^{\dagger} q_L + q_L^{\dagger} q_R  =\frac{1}{2}(\Psi^T \rho_2 t_2 \sigma^2 \Psi + \Psi^{\dagger} \rho_2 t_2 \sigma^2  (\Psi^{\dagger})^T) \nonumber \\
&&= \eta^T \big(t^2 \sigma^2 \rho^2 \mu^3 \big)\eta. \nonumber
\end{eqnarray}
%

The kinetic Hamiltonian with the Majorana representation is 
\begin{eqnarray}
&&\mathcal{H}_0 \equiv \eta^T i \gamma^k \partial_k \eta \nonumber \\
&& = \eta^T  i (\sigma^1 \partial_1+\sigma^2 \mu^2 \partial_2+\sigma^3 \partial_3)\eta .
\end{eqnarray}
For the reference, we summarize the Pauli matrix symbols,
\begin{eqnarray}
&&t^i \,:\, {\rm color}, \quad \sigma^i \,:\, {\rm Lorentz}, \quad \rho^i \,:\, {(q, \, \tilde{q})}, \nonumber \\
&&\mu^i \,:\, {\rm Majorana}, \quad \tau^i \,:\, {\rm flavor}
\end{eqnarray}
and the matrices of the kinetic term are
\begin{eqnarray}
\gamma^1 = t^0 \sigma^1 \rho^0 \mu^0 \tau^0 \quad \quad \gamma^2 = t^0 \sigma^2 \rho^0 \mu^2 \tau^0 \quad \quad \gamma^1 = t^0 \sigma^3 \rho^0 \mu^0 \tau^0 \nonumber
\end{eqnarray}
All zero components of the matrices are the identity matrices in the corresponding spinor spaces. 

Three more mass terms can be obtained in the flavor triplet channel. 
\begin{eqnarray}
i\, q_L^{\dagger}  \tau^j q_R - i\, q_R^{\dagger}  \tau^j q_L, \quad (j=1,2,3).
\end{eqnarray}
In the Majorana representation, the mass terms are
\begin{eqnarray}
\eta^T\big( t^2 \sigma^2 \rho^2 \mu^1 \tau^1 \big) \eta, \quad \eta^T \big(t^2 \sigma^2 \rho^1 \mu^3 \tau^2\big) \eta, \quad \eta^T \big( t^2 \sigma^2 \rho^2 \mu^1 \tau^3 \big) \eta. \nonumber
\end{eqnarray}
Two more mass terms are from superconducting channels
\begin{eqnarray}
&&\Big (q_L^T t^2 \sigma^2 \tau^2 q_L + q_L^{\dagger} t^2 \sigma^2 \tau^2 (q_L^{\dagger})^T \Big) + \Big( L \rightarrow R \Big), \nonumber \\
&&i \Big (q_L^T t^2 \sigma^2 \tau^2 q_L - q_L^{\dagger} t^2 \sigma^2 \tau^2 (q_L^{\dagger})^T \Big) -  \Big( L \rightarrow R \Big), \nonumber 
\end{eqnarray}
and in the Majorana representation they are 
\begin{eqnarray}
\eta^T \big(  t^2 \sigma^2 \rho^3 \mu^3 \tau^2 \big), \quad \eta^T \big( t^2 \sigma^2 \rho^0 \mu^1 \tau^2 \big) \eta. \nonumber
\end{eqnarray}
Thus, we have six mass terms, $\eta^T \Gamma^{\alpha} \eta$ with $\alpha=1,\cdots, 6$,
\begin{eqnarray}
&&\Gamma^1 = t^2 \sigma^2 \rho^2 \mu^3 \tau^0, \quad \Gamma^2 = t^2 \sigma^2 \rho^2 \mu^1 \tau^1, \quad \Gamma^3 = t^2 \sigma^2 \rho^1 \mu^3 \tau^2,  \nonumber \\
&& \Gamma^4 = t^2 \sigma^2 \rho^2 \mu^1 \tau^3, \quad \Gamma^5 =  t^2 \sigma^2 \rho^3 \mu^3 \tau^2, \quad \Gamma^6 = t^2 \sigma^2 \rho^0 \mu^1 \tau^2. \nonumber
\end{eqnarray}
Note that all gamma matrices $(\gamma^k, \Gamma^{\alpha})$ anti-commute each other (Clifford algebra) and all six order parameters are color-singlet. Thus, our construction is independent of introducing the Higg's doublet to control the gauge coupling strength.

The mass term Hamiltonian is parameterized by six fields,
\begin{eqnarray}
\mathcal{H}_m = g \, \eta^T {\phi}^{\alpha} \Gamma^{\alpha} \eta,
\end{eqnarray}
with a dimensionful coupling constant $g$. 
The six fields are matched with the $SU(4)$ generators from which the symmetry breaking down to $SP(4)$ is described by 
\begin{eqnarray}
\frac{SU(4)}{Sp(4)} \simeq S^5 \rightarrow \phi^{\alpha} \phi^{\alpha}=1.  
\end{eqnarray}
The last condition guarantees the correct ground state manifold.

\section{Anomaly coefficients and non-minimal models}

One useful diagnosis of the presence of anomalies is provided by 't Hooft.\citep{matching} 
Anomaly appears when a global symmetry is gauged and its gauge transformation produces physical effects, which are inconsistent with definition of gauge transformations. 
Such inconsistency with gauge transformations must be canceled because infinitesimally weak gauge coupling is always conceivable. 
To cancel it, one can add non-interacting massless fermions (so-called spectators) with opposite anomalies. 
Then, after spontaneous symmetry breaking or confinement, the original sector should contain some massless degrees of freedom to cancel the anomalies of the non-interacting fermions.  
 
Below, we follow the previous discussion in $SU(4)$ representation to consider the anomalies in different irreducible representations. \cite{su4}
A representation $R$ of the SU(4) group is characterized by three
numbers $q_1$, $q_2$, $q_3$, which are the numbers of columns in the
Young tableau with 1, 2, and 3 rows.

The dimension of the representation is
\begin{eqnarray}
D(R)=&& \frac1{12}(q_1+1)(q_2+1)(q_3+1)  \nonumber \\ 
&&\times(q_1+q_2+2)(q_2+q_3+2)(q_1+q_2+q_3+3) \nonumber
\end{eqnarray}
The anomaly is characterized by the number $A(R)$,
\begin{equation}
  \Tr (\{T^a_R,\, T^b_R\}T^c_R) = A(R) \Tr(\{t^a,\,t^b\}t^c)
\end{equation}
where $t^a$ refers to the fundamental representation of SU(4).
\begin{equation}
  A(R)=\frac1{60}(q_1-q_3)(q_1+q_3+2)(q_1+2q_2+q_3+4) D(R) \nonumber
\end{equation}
If $R=(q_1,q_2,q_3)$ then $\bar R=(q_3,q_2,q_1)$ and $A(\bar
R)=-A(R)$.  For real representations $q_1=q_3$ and $A(R)=0$.  In the
table below we exclude the complex conjugate representations.

\begin{tabular}{|c|c|c|c|c|c|c|}
\hline
$R$ &  \yng(1,1) & \yng(2) & \yng(2,1,1) & \yng(3) & \yng(2,1)   \\
\hline
$D(R)$ &  6 & 10 & 15 & 20 & 20  \\
\hline
$A(R)$ &  0 & 8 & 0 & 35 & 7 \\
\hline
\hline
$R$ & \yng(2,2) &\yng(4) & \yng(3,1,1) & \yng(3,1) & \yng(3,3) \\
\hline
$D(R)$ & 20 & 35 & 36 & 45 & 50\\
\hline
$A(R)$ & 0 & 112 & 21 & 48 & 0\\
\hline
\end{tabular}
with $D\Big(\yng(1)\Big)=4$ and $A\Big(\yng(1)\Big)=1$.

A class of non minimal models (weakly coupled one) can be constructed by two conditions : 1) anomaly matching and 2) presence of Yukawa coupling (fermion mass in the symmetry broken phase). 
The formal expression for the anomaly condition is 
\begin{eqnarray}
\sum_{R} n_R \mathcal{A}(R) = k, 
\end{eqnarray}
$k$ is the WZW level and the summation is over all representations. 
Possibility of the Yukawa coupling can be obtained by multiplication of representations. 
%

\section{Differential Geometry}

The interior derivative is defined as
\begin{eqnarray}
i_a \Omega = \frac{1}{(n-1)!} \xi_a^j \Omega_{j i_2 \cdots i_n} d\phi^{i_2} \cdots d\phi^{i_n}
\end{eqnarray}
with a vector field $\xi_a = \phi^i (t_a)_{ij} \partial_j$.

Since the volume form is the highest form, we have nice properties
\begin{eqnarray}
&&d\omega=0, \quad i_a \omega = d v_a, \quad \mathcal{L}_a(\omega)=0 \nonumber 
\end{eqnarray}
We introduce $\widehat{\Omega}$ notation for replacing a partial derivative with a covariant derivative ($\Omega(\partial) \rightarrow \Omega(D) $). 

For $1+1$ dimensions, one can show that
\begin{eqnarray}
i_a(\omega) &=& \frac{1}{3!} \epsilon_{ijkl} \phi^i \, i_a(d\phi^j) d\phi^k d\phi^l+\cdots \nonumber \\
&=&  \frac{1}{2 \cdot 2!} \varepsilon_{ijkl} (t_a)_{ij} d\phi^k d \phi^l= d v_a\nonumber
\end{eqnarray}
 which defined the one form, 
\begin{eqnarray}
v_a = \frac{1}{2 \cdot 2!} \varepsilon_{ijkl} (t_a)_{ij} \phi^k d \phi^l. \nonumber
\end{eqnarray}  
The relation $(t_a)_{ij} = \frac{1}{2} \epsilon_{ijkl} (\omega_a)_{kl}$ is useful.
One can show $\mathcal{L}_a v_b = f_{ab}^c v_c$ straightforwardly. 

The interior derivative of the one form is 
\begin{eqnarray}
i_b (v_a) = \frac{1}{2}\epsilon_{a_1 a_2 b_1 b_2} \Big( (\phi^{b_1})^2 + (\phi^{b_2})^2  \Big).
\end{eqnarray}
Its symmetrized one is 
\begin{eqnarray}
\frac{i_a (v_b)+i_b (v_a)}{2} = \frac{1}{4}\epsilon_{a_1 a_2 b_1 b_2}
\end{eqnarray}
 
One useful identity  is 
\begin{eqnarray}
d \widehat{\Omega} = \widehat{d\Omega} - F^a \widehat{i_a \Omega}+ A^a \widehat{ \mathcal{L}_a \Omega }. \nonumber 
\end{eqnarray}
One can show this by using components of the form. 
By using the identity twices, one can obtain 
\begin{eqnarray}
d \widehat{\omega} &=& -F^a \widehat{d v_a} =-F^a (d \widehat{v_a} +F^b i_{b} v_{a} - A^b \mathcal{L}_b v_a) \nonumber\\
&=& - d(F^a \widehat{v}_a)- F^a F^b i_{(b} v_{a)}. \nonumber
\end{eqnarray}
Thus, the anomaly coefficient is 
\begin{eqnarray}
d_{ab} = \frac{i_a (v_b)+i_b (v_a)}{2} = \frac{1}{4}\epsilon_{a_1 a_2 b_1 b_2}
\end{eqnarray}

For $3+1$ dimensions, one can show that
\begin{eqnarray}
v_a = \frac{1}{2 \cdot 4!} \varepsilon_{ijklmn} (t_a)_{ij} \phi^k d \phi^l d \phi^l d\phi^m, \quad \mathcal{L}_a v_b = f_{ab}^c v_c. \nonumber
\end{eqnarray}
By using the identity, we obtain 
\begin{eqnarray}
d \widehat{\omega} 
&=& -d (F^a \widehat{v_a}) -F^a F^b \widehat{d v_{(ab)}}  \nonumber \\
&=& -d (F_a \widehat{v_a} +F^a F^b \widehat{ v_{(ab)}} )- d_{abc} F^a F^b F^c. \nonumber
\end{eqnarray}
The relation $i_{(a} v_{b)} = d v_{(ab)}$ is used, and the anomaly coefficient is 
\begin{eqnarray}
d_{abc} =  \frac{i_c  v_{(ab)} +i_b  v_{(ac)}+i_a  v_{(cb)} }{3}= \frac{\epsilon_{a_1 a_2 b_1 b_2 c_1 c_2}}{4!}.
\end{eqnarray}

\section{WZW term from integrating out fermions}
Main purpose of this section is to provide explicit informaiton on fermion determinant calculation. 

In $1+1$ dimensions, let us consider the four component complex spinors, $\Psi$, with two pauli matrices ($\sigma^i, \tau^j$). 
The action is 
\begin{eqnarray}
S &=& \int \Psi^{\dagger} (\partial_{\tau} - i \alpha \sigma^z \tau^0 \partial_x)\Psi + g \phi^i \Psi^{\dagger} M_i \Psi \nonumber
\end{eqnarray}
with anti-commuting matrices $\{M_i, M_j\}=2 \delta_{ij}$ and $ \{\sigma^z, M_j\}=0 $. 
The coupling constant $\alpha=\pm1$ will determines the sign of the WZW term. 
In the next senction, the mass matrices ($M_i$) are explicitly constructed. 
Since the matrix size is $4 \times 4$, the total number of the matrices is five. 

The effective aciton is defined 
\begin{eqnarray}
\Gamma^{eff}= -\log \mathcal{Z}, \quad  \mathcal{Z}[\phi] = \int_{\psi, \psi^{\dagger}} \exp{(-S_f)} = {\rm Det} \mathcal{D} \nonumber
\end{eqnarray}
with $\mathcal{D}=\partial_{\tau} -i \alpha  \sigma^z \tau^0 \partial_x+g\hat{\phi}$ and $ \hat{\phi}= \phi^i M_i $.

The variation of the effective action is 
\begin{eqnarray}
-\delta_{\phi}\Gamma^{eff}_2 &=& {\rm tr}(\delta D D^{\dagger}(DD^\dagger)^{-1}) \nonumber \\
&=&  g^2 {\rm tr} (\delta \hat{\phi} \cdot \hat{\phi} \,G_0 (1+M G_0 +\cdots) ). \nonumber
\end{eqnarray}
The Green's funciton is introduced by the operator, 
\begin{eqnarray}
D D^{\dagger} = -\partial_{\tau}^2 - \partial_x^2+g^2 +g(\partial_{\tau} \hat{\phi}- i \alpha \sigma^z \partial_x \hat{\phi}) \equiv G_0^{-1} -M. \nonumber
\end{eqnarray}
$G_0^{-1}(k, \omega) = \omega^2 +k^2+g^2$.
With this set-up, we do the gradient expansion in terms of $g$ (or $M$).

In $1+1$ dimensions, the topological term can be obtained by the second term
\begin{eqnarray}
-\delta_{\phi}\Gamma^{(2)}_2
&=& g^2 {\rm tr}(\delta\hat{\phi} \cdot \hat{\phi} M^2 G_0^3) \nonumber \\
&=& i  \int_x \frac{\alpha}{2\pi} \varepsilon_{\mu \nu} \varepsilon_{ijkl} \delta \phi^i \phi^j \partial_{\mu} \phi^k \partial_{\nu} \phi^l. \nonumber
\end{eqnarray}
The relation $ {\rm tr}(M_i M_j M_k M_l \sigma^z) =4 \varepsilon_{ijkl}$ is used. 
From this calculation, one can further notice that the sign of the imaginery term depends on the sign structure of the Hamiltonian. 
Instead of $\mathcal{H}_0 =\Psi^{\dagger} ( - i \sigma^z \tau^0 \partial_x)\Psi  $

In $3+1$ dimensions, let us consider the 16 component complex spinors, $\Psi$, with four pauli matrices. 
There are nine gamma matrices for  the Clifford algebra which are construced in the previous section. 
The $N_c=2$ QCD with $N_f=2$ theory has the same number of degrees of freedom. 
Following the previous calculation, we introduce
\begin{eqnarray}
S &=& \int \Psi^{\dagger} (\partial_{\tau} -i \alpha \gamma^{s}\partial_{s})\Psi + g \phi^i \Psi^{\dagger} \Gamma_i \Psi \nonumber \\
-\Gamma^{eff}&\equiv& \log D =\log (\partial_{\tau} -i \alpha \gamma^{r}\partial_{r}+g\hat{\phi}), \quad  \hat{\phi}= \phi^i \Gamma_i \nonumber
\end{eqnarray}
with $s=1,2,3$ for the gamma matrices in the kinetic term and $i,j=1,\cdots 6$ for the gamma matrices in the mass terms
($ \{\Gamma_{i}, \Gamma_{j}\}=2 \delta_{ij}$).
The variation of the effective action is 
\begin{eqnarray}
-\delta_{\phi}\Gamma^{eff}  &=& {\rm tr}(\delta D D^{\dagger}(DD^\dagger)^{-1}) \nonumber \\
&=&  g^2 {\rm tr} (\delta \hat{\phi} \cdot \hat{\phi} \,G_0 (1+M G_0 +\cdots) ) \nonumber
\end{eqnarray}
with 
\begin{eqnarray}
D D^{\dagger} = -\partial_{\tau}^2 - \partial^2+g^2 +g(\partial_{\tau} \hat{\phi}- i \alpha\gamma^r \partial_r \hat{\phi}) \equiv G_0^{-1} -M. \nonumber
\end{eqnarray}
In $3+1$ dimensions, the topological term can be obtained by the fourth term
\begin{eqnarray}
-\delta_{\phi}\Gamma^{(4)} &=& g^2 {\rm tr}(\delta\hat{\phi} \cdot \hat{\phi} M^4 G_0^5) \nonumber \\
&=& i  \alpha \int_x  \frac{\varepsilon_{\mu \nu \rho \lambda} \varepsilon_{ijklmn}}{12\pi^2} \delta \phi^i \phi^j \partial_{\mu} \phi^k \partial_{\nu} \phi^l \partial
_{\rho} \phi^m \partial_{\lambda} \phi^n  \nonumber
\end{eqnarray}
, which is the exactly same as the variation of the WZW term.

\section{Group theory consideration of anomalies }
In this section, we provide more information on group theoretical relations in non-abelian anomalies. The group theoretical relation is especially powerful to find anomalies in fermion only theories. One can calculate either  a loop-diagram or the Fujikawa measure in path integral. \cite{weinberg,harvey}    

\subsection{1+1 dim : $SO(4) \sim SU(2)_{L} \times SU(2)_R$}
We use the Clifford algebra to connect representations of $SO(4)$ group and $ SU(2)_{L} \times SU(2)_R$. 
The idea is $SO(5)$ group has a natural spinor representation in terms of five Gamma matrices, and we use four gamma matrices out of the five. The remaining one becomes the ``chiral'' operator. 
To construct it, we use
\begin{eqnarray}
&&\Gamma^5 = \sigma^3 \tau^0 \nonumber \\
&&\Gamma^1 = \sigma^2 \tau^1, \quad \Gamma^2 = \sigma^2 \tau^2, \quad \Gamma^3= \sigma^2 \tau^3, \quad \Gamma^4 = \sigma^1 \tau^0\nonumber
\end{eqnarray}
It is straightforward to relate them to the Lie-algebra of $SO(4)$. We define $M_{ij}=\frac{1}{2} \Gamma^i \Gamma^j$. 
Explicitly, we have 
\begin{eqnarray}
&&M_{12} = \frac{i}{2} \sigma^0 \tau^3, \quad M_{13}=\frac{i}{2} \sigma^0 \tau^2, \quad M_{23}=\frac{i}{2} \sigma^0 \tau^1 \nonumber \\
&&M_{14} = \frac{i}{2} \sigma^3 \tau^1=\Gamma^5 M_{23}, \quad M_{24}=\frac{i}{2} \sigma^3 \tau^2=\Gamma^5 M_{13} \nonumber \\
&&M_{34}=\frac{i}{2} \sigma^3 \tau^1 = \Gamma^5 M_{12} \nonumber
\end{eqnarray}
Note that $\Gamma^5$ indeed plays a role as the chiral operator. 

The anomaly coefficient is related to
\begin{eqnarray}
&&\tilde{d}_{ab} =  {\rm Tr}(\Gamma^5 M_{a} M_{b}) = \varepsilon_{a_1 a_2 b_1 b_2}. \nonumber 
\end{eqnarray}

\subsection{3+1 dim : $SO(6) \sim SU(4)$}
As in $1+1$ dimensions, we can use the $SO(7)$ representations. Seven gamma matrices are necessary, and their minimum size is $8 \times 8$. One needs three types of Pauli matrices and we use the notation $(\sigma^{\alpha \beta \gamma} = \sigma^{\alpha} \otimes \sigma^{\beta} \otimes \sigma^{\gamma})$ with Pauli matrices $\sigma^{\alpha}$, $\alpha =0,1,2,3$.  $\sigma^0$ is the $2 \times 2$ identity matrix. Similarly,  $(\sigma^{\alpha \beta } = \sigma^{\alpha} \otimes \sigma^{\beta})$.

From the $SO(7)$ algebras, we pick one operator as a projection operator.   
\begin{eqnarray}
&&\Gamma^7 = \sigma^{300} \nonumber \\
&&\Gamma^6 = \sigma^{210}, \quad \Gamma^5 = \sigma^{220}, \quad \Gamma^4= \sigma^{230}, \nonumber \\
&&\Gamma^3 = \sigma^{103}, \quad \Gamma^2 = \sigma^{102}, \quad \Gamma^4= \sigma^{101}, \nonumber
\end{eqnarray}
With $M_{ij}= \Gamma^i \Gamma^j$, we have 
\begin{eqnarray}
&&M_{12} = i \sigma^{003}, \quad M_{13}=i \sigma^{002}, \quad M_{14}=i \sigma^{331}, \quad M_{15}=i \sigma^{321}, \nonumber \\
&&M_{16}=i \sigma^{311}, \quad M_{23}=i \sigma^{001}, \quad M_{24}=i \sigma^{332}, \quad M_{25}= i \sigma^{322}, \nonumber \\
&& M_{26}=i \sigma^{312}, \quad M_{34}=i \sigma^{333}, \quad M_{35}=i \sigma^{323}, \quad M_{36}=i \sigma^{313}, \nonumber \\
&& M_{45}=i \sigma^{010}, \quad M_{46}=i \sigma^{020}, \quad M_{56}=i \sigma^{030} \nonumber 
\end{eqnarray}
Note that $\Gamma^7$ indeed plays the chiral operator. 
Thus, we can choose either $+1$ or $-1$ component and obtain the spinor representation of $SO(6)$. 
\begin{eqnarray}
&&M_{12} =   \sigma^{03}, \quad M_{13}=  \sigma^{02}, \quad M_{14}=  \sigma^{31}, \quad M_{15}=  \sigma^{21}, \nonumber \\
&&M_{16}=  \sigma^{11}, \quad M_{23}=  \sigma^{01}, \quad M_{24}=  \sigma^{32}, \quad M_{25}=   \sigma^{22} \nonumber \\
&& M_{26}=  \sigma^{12}, \quad M_{34}=  \sigma^{33}, \quad M_{35}=  \sigma^{23}, \quad M_{36}=  \sigma^{13} \nonumber \\
&& M_{45}=  \sigma^{10}, \quad M_{46}=  \sigma^{20}, \quad M_{56}=  \sigma^{30}. \nonumber 
\end{eqnarray}
These are also $SU(4)$ fundamental representations. 
The anomaly coefficient is proportional to 
\begin{eqnarray}
\tilde{d}_{abc}= {\rm Tr}(M_a \{M_b, M_c \})= \varepsilon_{a_1 a_2 b_1 b_2 c_1 c_2}  \nonumber
\end{eqnarray}
upto a non-zero numerical constant.

\end{document}